\documentclass[9pt,english]{article}
\usepackage[a4paper,left=2cm,top=2cm,right=2cm,bottom=2.5cm]{geometry}
\usepackage[utf8]{inputenc} 
\usepackage[T1]{fontenc} 
\usepackage{babel} 
\usepackage{csquotes} 
\usepackage{xcolor} 
\usepackage{graphicx}

\usepackage{palatino} 

\usepackage{setspace} 

\usepackage{scrpage2} 
\ohead{}
\ihead{}
\ofoot{\pagemark}
\ifoot{}
\let\OLDthebibliography\thebibliography
\renewcommand\thebibliography[1]{
  \OLDthebibliography{#1}
  \setlength{\parskip}{0pt}
  \setlength{\itemsep}{0pt plus 0.3ex}
}

\begin{document}

\begin{center}
{\bf Comment on arXiv:1505.07451}

Y. Lu, E. Benckiser, B. Keimer

Max Planck Institute for Solid State Research, Stuttgart, Germany
\end{center}

\onehalfspacing

The ground state of bulk nickelates with composition $R$NiO$_3$ is either charge-disordered and paramagnetic (for $R =$ La), or charge-ordered and antiferromagnetic (for $R =$ Pr, Nd, and other rare earths). \cite{Torrance} Raman scattering experiments on PrNiO$_3$-PrAlO$_3$ superlattices \cite{Hepting} have recently revealed a state with antiferromagnetic order but no (or at least greatly suppressed) charge order, in agreement with predictions resulting from theoretical work. \cite{Balents} In a preprint that does not discuss Ref. \cite{Hepting}, Meyers {\it et al.} \cite{Meyers} claim to have observed a closely related state in a 15-unit-cell thin film of NdNiO$_3$. In this brief note, we show that the evidence presented by Meyers {\it et al.} is insufficient to support this claim.

The experimental protocol used by Meyers {\it et al.} follows earlier work by Staub {\it et al.} \cite{Staub} who used resonant x-ray scattering (RXS) to detect charge order in $\sim 130$ unit-cell-thick films of NdNiO$_3$. However, the much lower signal intensity combined with interference from the substrate complicate RXS experiments on ``ultrathin'' films. Specifically, Meyers {\it et al.} use the apparent absence of a resonant enhancement of the (105) reflection at the Ni K-absorption edge to argue that charge order is absent in their film. They then use the photon energy ($E$) dependence of the (220) reflection to argue that resonance effects at (105) would have been detectable if they had been present. This line of argument is fallacious, not only because the intensity of (220) is four orders of magnitude larger (and the data quality correspondingly higher) than the one of (105), but also because the $E$-dependences of the (220) and (105) reflections arise from different mechanisms.

This becomes obvious if one writes out the structure factor of the rock-salt-like charge order, which comprises Ni sites in two sublattices of the pseudocubic perovskite structure, {\it i.e.} Ni$_1$ at (0.0,0.0,0.0) and (0.5,0.5,0.5), and Ni$_2$ at (0.0,0.0,0.5) and (0.5,0.5,0.0):

\[
F_{220}(q, E) = A_{O,Nd}(q)+2f_1(q,E)+2f_2(q,E)
\]
and
\[
F_{105}(q, E) = B_{O,Nd}(q)+2f_1(q,E)-2f_2(q,E)
\]

where $f$ is the Ni form factor. Each $f$ can be written as

\[
f(q,E)=f_0(q)+f'(E)+i f''(E)
\]

where $f_0(q)$ is the conventional (Thomson) part and $f'$, $f''$ the resonant correction. Energy scans at (220) probe the {\it sum} of the form factors of the two Ni sublattices, and the $E$-dependence of the intensity merely reflects the shape of the real part of the Ni form factor, $f'$ (upper panel of Fig. 4 in Ref. \cite{Staub}). The $E$-dependence of the (105) intensity, on the other hand, probes the {\it difference} of the two Ni sublattices. So the fact that one can see the former in no way implies that one could see the latter as well, even if charge order were present.

In addition, the structure of NdNiO$_3$ \cite{Garcia} implies that $A_{O,Nd} >> B_{O,Nd}$; this is responsible for the four-orders-of-magnitude larger intensity of (220) relative to (105). The total intensity reads

\[
I = (A/B+2f_1'\pm 2f_2')^2+ (2f_1''\pm2f_2'')^2
\]

with $(A, +)$ for (220) and $(B, -)$ for (105). We see that the amplitude of the $E$-dependent contribution is proportional to $A$ and $B$ for the (220) and (105) reflections, respectively. This again greatly facilitates detection of the $E$-dependence of the (220) intensity relative to (105).

Meyers {\it et al.} cite the apparent absence of the (015) reflection as an additional argument for a charge-disordered state in their film. \cite{Meyers} However, if one calculates the intensities of (105) and (015) based on the structure of bulk NdNiO$_3$ \cite{Garcia}, the former turns out to be about $\sim 1000$ times larger than the latter in the relevant $E$-range. In Staub's work \cite{Staub}, the ratio of the intensities at resonance is about 10 (perhaps due to partial twinning of their film). In Fig. 4A of Meyers {\it et al.}, a substrate reflection dominates the intensity at the position of the main (015) reflection, and even a factor-of-ten reduction relative to (105) would bring the intensity at the first Kiessig fringe of (015) down to the noise level.

In closing, we emphasize that our arguments do not necessarily imply that the conclusion stated by Meyers {\it et al.} is incorrect, but they do illustrate possible pitfalls which we hope can be avoided in future RXS studies of thin films and heterostructures.

\begingroup
\renewcommand{\section}[2]{}%

\endgroup

\end{document}